\documentclass[12pt]{iopart}

\usepackage{citesort}

\usepackage{soul}
\usepackage{color}
\usepackage{graphicx}% Include figure files
\begin{document}

\title[Stimulated Smith-Purcell emission]{Stimulated Smith-Purcell emission based on bound states in the continuum}

\author{Zhaofu Chen$^1$, Renjun Yang$^2$ and Xiaohan Sun$^1$}
\address{$^1$Research Center for Electronic Device and System Reliability, Southeast University, Nanjing 210096, China}%
%\ead{xhsun@seu.edu.cn}
\address{$^2$High Energy Accelerator Research Organization~(KEK), Tsukuba 305-0801, Japan}%
% kek email address: renjun.yang@kek.jp

\ead{chen.zhaofu@outlook.com}

%\author{Content \& Services Team}
%
%\address{IOP Publishing, Temple Circus, Temple Way, Bristol BS1 6HG, UK}
%\ead{submissions@iop.org}
\vspace{10pt}
\begin{indented}
\item[]December 2021
\end{indented}

%\linenumbers
\begin{abstract}
Recent advances in the development of bound states in the continuum offer new strategies to tailor electron-wave interaction and hence control the electron-induced emission. In this article we investigate the design to produce stimulated Smith-Purcell emission with a single open grating. This scheme exploits a strong radiative resonance near a bound state in the continuum, enabling staggering enhancements of multiple diffraction orders of a subwavelength grating under evanescent wave incidence. The interaction between a continuous electron beam and the radiative resonant mode bunches electrons, resulting in coherent oscillation and consequently stimulated Smith-Purcell radiation. Using a higher diffraction order for Smith-Purcell radiation, coherent radiation with low-energy electrons is also allowed. The interaction with a radiative mode that has two propagating space harmonics enables stimulated radiation towards two different directions. This work paves the way to a compact coherent radiation source, which may find application in communications, physics, and biology.
\end{abstract}

%
% Uncomment for keywords
%\vspace{2pc}

%\vspace{2pc}
% Uncomment for Submitted to journal title message
%\submitto{\NJP}
%
% Uncomment if a separate title page is required
%\maketitle
% 
% For two-column output uncomment the next line and choose [10pt] rather than [12pt] in the \documentclass declaration
%\ioptwocol
%

\section{Introduction} \label{sec:introduction}

In the past decade, growing interest in compact radiation sources has led to extensive analysis of various interactions between free electrons and surrounding media \cite{PhysRevB.103.075403,Pupasov_Maksimov_2021,Lin2018,Liu2017e,https://doi.org/10.1002/lpor.202000149,PhysRevLett.116.205003,PhysRevLett.123.057402,doi:10.1021/acsphotonics.9b00251,Roques-Carmes2019,Wang2016,Gardelle2010,Aryshev2020,Urata1998}. Among them, Smith-Purcell (SP) radiation emitted by electrons traveling over a periodic structure is an important method for realizing free-electron radiation sources~\cite{PhysRevLett.123.060401,PhysRevX.7.011003}. Ordinary SP radiation in the spontaneous emission regime has been observed from microwave to optics~\cite{PhysRevLett.74.3808,Ye:19}. However, such radiation is typically featured by incoherence and low power, although many mechanisms have been proposed to improve the efficiency \cite{Yang2018,Liu2014,Song2018}, e.g., coupling of electrons with bound states in the continuum (BICs). Methods to improve the radiation power are of great necessity for its applications in communications, physics, and biology.

The existing approaches for coherently enhancing the SP radiation fall into two major categories: Orotrons and Smith-Purcell free-electron lasers (SP-FELs). The Orotron produces stimulated emission by utilizing the interaction between a continuous electron beam and a self-excited standing wave in an external Fabry-Perot (FP) cavity~\cite{doi:10.1063/1.3079678,Liu2016,Grishin2004,Myasin2016}. Commercially available Orotrons can provide continuous-wave radiation with a watt level of output power at millimeter wavelengths~\cite{Bratman2010}. However, the complicated cavity structure tends to hinder its wide application. The SP-FEL generates superradiant SP radiation with an open grating. It uses a surface mode bound at the grating to bunch a continuous electron beam, and the arising electron bunches generate coherent spontaneous SP radiation at the harmonics of the surface mode frequency \cite{PhysRevA.40.876,Andrews2005a,Andrews2004,Andrews_2006,Kumar2006,PhysRevE.57.1061}. The simple structure and the described capability of frequency multiplication of SP-FELs have drawn considerable interest. However, the efficiency and radiation power of SP-FELs are limited by the spontaneous nature of the radiation~\cite{doi:10.1063/1.3554435}.
%in simulations and experiments Methods to improve the efficiency of SP-FELs usually involve using two cascade gratings, but at the cost of complexity \cite{Prokop2010,Liang2017,Liu2018a}. %or introducing extra cavities \cite{doi:10.1063/1.3554435,zhou2016electron}

%\sout{Here, we present a method in which the strong resonance in a single dielectric SWG is used to achieve passive bunching in tandem with the enhancement of the coherent SP radiation at the bunching frequency.}
In this article, we investigate the high-power stimulated SP radiation enabled by BICs. In contrast to the Orotron employing a two-mirror cavity, an open grating is employed for the electrodynamic system. Theoretical analysis of then grating upon evanescent wave incidence will be performed. Numerical investigations of the interaction between a continuous electron beam and the grating will be presented, followed by a discussion over the applicability to lower electron energies and different resonant modes. These analyses demonstrate that this method could produce coherent SP radiation efficiently, potentially enabling a compact terahertz source. %which leads to physical insights into the BICs excited by the nonradiative incident wave bound at the electron beam. We show that the interaction between a continuous electron beam with a resonant radiative mode near a BIC can lead to coherent oscillation, and consequently high-power coherent SP radiation. We also show that the required electron energy can be reduced by using a higher diffraction order for SP radiation. Our method provides an efficient way to produce coherent SP radiation, potentially enabling a compact terahertz source. 

%In Fig. 4,5 the authors present a spectrum of the resulting emission. Is this indeed the spectrum that is numerically calculated? If so, it is in a very narrow frequency range, obscuring the possibility to view energy transfer to other interaction orders. I suggest that the authors add a simulation in which they verify that the electron bunching and efficiency of their radiation source remain similar even when the simulation includes the possibility to emit to other resonant modes (which is what will happen in a practical experiment).\\

\section{Model description} \label{sec:model}

\begin{figure}[htbp!]
	\centering
	\includegraphics[width=11cm]{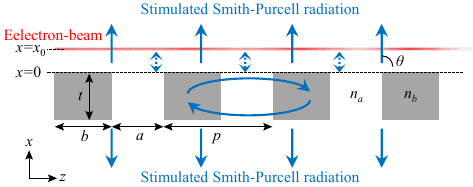}
	\caption{A schematic drawing of a dielectric grating interacting with an electron beam. Electrons (red) travel in the $z$ direction with an initial velocity of $v_{\mathrm{0}}$. $n_{\mathrm{b}}$ and $n_{\mathrm{a}}$ are the refractive indices of the high-index and low-index bar, respectively, %. $x=0$ is located at the grating upper plate. Parameters: 
	$p$ the grating period, $t$ the grating thickness, $b$ the high-index bar width, $a$ the low-index bar width, $\theta$ the radiation angle. The distance between the beam center and grating is $x_0$. }% (b) The longitudinal electronic field of the evanescent 0th diffraction order (up) and the -1st diffraction order (down), with the -1st-order being radiative at $\theta=90^\circ$. (c) Electron energy gain with respect to the start phase arising from the interaction with the 0th diffraction order.} 
	\label{fig:schematic}
\end{figure}

Figure~\ref{fig:schematic} schematically illustrates the concept of generating stimulated SP radiation with a single open grating, which supports a strongly resonant radiative mode near a BIC. When an electron beam moves over, the radiative mode is excited by the evanescent incident field. An evanescent space harmonic of the radiative mode will in turn modulate the electron energy and intensity, leading to an enhancement of the incident field at the frequency of the radiative mode. Consequently, effective interaction between the electrons and the radiative mode takes place, resulting in coherent oscillation that transfers the electron energy to the radiative mode. The SP radiation, corresponding to those propagating space harmonics of the radiative mode, can be significantly enhanced due to the beam-wave interaction.

The associated mechanism is described in detail as follows. For simplicity, an infinite extent in the $y$ direction of the structure depicted in figure~\ref{fig:schematic} is assumed. The dielectric grating comprises of alternating high-index ($n_{\rm{b}}$) and low-index ($n_{\rm{a}}$) bars of thickness $t$. $x = 0$ is located at the top of the grating. The electron beam can be treated as a number of line charges moving in close proximity to the grating at $x = x_0$ with a charge density distribution $q$ per unit length in the $y$ direction and an initial velocity of $v_0=\beta_0 c$ in the $z$ direction, with $c$ being the speed of light. The current density in the time domain can be written as~\cite{Kumar2006,vandenBerg:73,Yang2018}
\begin{equation}
	{\bi J}(x,z,t) = 
	\widehat{\bi{z}}qv_0\delta(x-x_0)\sum_{j=1}^M\delta\left[z-v_0t-\Delta z_j\right],
	\label{eq:current_time}
\end{equation}
where $M$ is the total number of line charges, $\Delta z_j$ is the position of the $j$th line charge at $t=0$. The current density in the frequency domain is then obtained via Fourier transform, as 
\begin{equation}
	{\bi J}(x,z,\omega) =  
	\widehat{\bi{z}}q\delta(x-x_0)\varsigma{\rm e}^{{\rm i}k_{z0}z},
	\label{eq:current}
\end{equation}
where $k_{z0} = \omega/v_0$ and $\varsigma=\sum_{j=1}^M{\rm e}^{{\rm -i}k_{z0}\Delta z_j}$ \cite{Kumar2006}. Between the beam and grating, the fields induced by the current can be characterized by its magnetic field:
\begin{equation}
	{\bi H}^{\rm in} = \widehat{\bi{y}}\frac{q}{2}\varsigma{\rm e}^{{\rm i}\left[k_{z0}z-k_{x0}(x-x_0)\right]},
	\label{eq:field_in}
\end{equation}
where $k_{x0} = {(k_0^2-k_{z0}^2)}^{1/2}$ is the $x$ wavenumber, $k_{0} = \omega/c$ the free-space wavenumber~\cite{vandenBerg:73,Yang2018}. With $\beta_0<1$, $k_{x0}$ is imaginary, and the field described by (\ref{eq:field_in}) is nonradiative. 

Upon evanescent wave incidence, the periodic nature of the grating suggests that the reflected and transmitted waves can be expanded into a series of space harmonics. Considering the ordinary SP radiation in the spontaneous emission regime, where the beam-wave interaction can be neglected, the fields of the $n$th diffraction order above the grating can be given by
\begin{eqnarray}
	&&{\bi H}^{\rm re}_{n}=\widehat{\bi{y}}\frac{r_nq}{2}\varsigma{\rm e}^{{\rm i}k_{x0}x_0}{\rm e}^{{\rm i}\left(k_{zn}z+k_{xn}x\right)},
	\label{eq:fieldH_re}\\
	&&{\bi E}^{\rm re}_{n}=\left[\widehat{\bi{x}}k_{zn}-\widehat{\bi{z}}k_{xn}\right] \frac{r_nq}{2\omega\varepsilon_0}\varsigma{\rm e}^{{\rm i}k_{x0}x_0}{\rm e}^{{\rm i}\left(k_{zn}z+k_{xn}x\right)},\qquad
	\label{eq:fieldE_re}
\end{eqnarray}
where $r_n$, $k_{zn} = k_{z0} + 2n\pi/p$, and $k_{xn} = {(k_0^2-k_{zn}^2)}^{1/2}$ are the reflection coefficient, $z$ wavenumber and $x$ wavenumber of the $n$th diffraction order, respectively~\cite{Kumar2006,vandenBerg:73,Yang2018}. 

The eigenmodes of the grating play an essential role in determining the diffraction properties. All the space harmonics are evanescent for a surface eigenmode, while one or more propagating space harmonics emerge for a radiative eigenmode. %The space harmonics with a real $k_{xn}$ can propagate into space, forming the SP radiation, and the rest with an imaginary $k_{xn}$ is evanescent in the $x$ direction. Note that only negative diffraction orders can propagate. 
The SP emission requires the excitation of radiative modes. %, e.g., the mode with wavelength $\lambda=p/\beta_0$ where only the -1st order is propagating. % and the other orders being evanescent. 
%The fields of the 0th and -1st orders are schematically shown in figure~\ref{fig:schematic}(b). 
In this case, rewriting the $z$ wavenumber $k_{zn}$ in terms of the radiation angle $\theta_n$ for those propagating space harmonics, $k_{zn} = k_0 \cos\theta_n$, one obtains the well-known relation between the radiation wavelength $\lambda$ and the electron velocity
\begin{equation}
	\lambda = -\frac{p}{n}(\frac{1}{\beta_0} - \cos\theta_n).
	%n\lambda = p(\cos\theta-\beta^{-1}).
	\label{eq:field_angle}
\end{equation}

To produce SP radiation efficiently, a considerable deal of work has been devoted to studying the evanescent-to-radiative wave conversion efficiency \cite{vandenBerg:74,PhysRevE.49.3340,PhysRevE.57.1075,Yang2018,Song2018}, while the coexisting evanescent space harmonics of a radiative mode have seldom been utilized. In contrast, we will show that, by exploiting a resonant radiative mode near a BIC, the 0th-order space harmonic can bunch electrons and enable stimulated SP radiation.
%For SP radiation sources with either metallic gratings or dielectric gratings, a large deal of work has been devoted to studying the evanescent-to-propagating wave conversion efficiency when a radiative mode is excited\cite{vandenBerg:74,PhysRevE.49.3340,PhysRevE.57.1075,Yang2018,Song2018}, while the coexisting evanescent modes have seldom been utilized. In contrast, by exploiting a dielectric grating with a strong resonance near a BIC, we use an evanescent diffraction mode to bunch electrons, enabling coherent SP radiation.
Apperently, the 0th-order space harmonic %has a $z$ wavenumber $k_{z0}$, indicating that it 
is phase synchronous with the electrons. %Inspired by the velocity bunching scheme in dielectric laser accelerators \cite{PhysRevLett.123.264803,PhysRevLett.121.214801,PhysRevLett.123.264802,RevModPhys861337}, The longitudinal electric field of the 0th-order \sout{spatial} {\color{red}{space}} harmonics can be utilized to accelerate or decelerate electrons. The energy gain of a phase-matched electron over one grating period can be expressed as 
%Using the longitudinal electric field of the 0th-order space harmonic to accelerate or decelerate electrons, 
The energy gain of a phase-matched electron from the longitudinal electric field over one grating period is
\begin{equation}
	\Delta E = {\rm i}k_{x0}|r_0|q(2\omega\varepsilon_0)^{-1}ep|\varsigma|{\rm e}^{{\rm i}k_{x0}\left(x_0+x\right)}\sin\omega t_0,
	\label{eq:energy}
\end{equation}
where $e$ is the elementary charge, $\omega t_0$ is the start phase of an electron. % relative to the longitudinal electric field of the 0th order.
%As shown in figure~\ref{fig:schematic}(c), 
Electrons with $\omega t_0 \in \left[0,\pi\right]$ lose energy, and electrons with $\omega t_0 \in \left[\pi,2\pi\right]$ gain energy. Such dependence leads to electron energy modulation, which bunches the electrons at the frequency of the radiative mode. %evolves into density modulation after a distance and thus generates a train of electron bunches. The fields experienced by electrons depend on the value of $|\varsigma|$, which is relevant to the beam distribution~\cite{Andrews2005a,RevModPhys.91.035003}. 
For periodically bunched beams, the incident fields induced by the electrons concentrate in the bunching frequency and its higher harmonics, thereby leading to coherent oscillation that transfers energy from the beam to the radiative mode~\cite{Andrews2005a}. %When the energy is transferred from the electron beam to the radiative mode due to the beam-wave interaction, 
%Owing to the energy transfer, the fields of the radiative mode are coherently enhanced. 
Note that all those space harmonics of a single eigenmode operate at the same frequency and are not independent~\cite{zhang1998electromagnetic}. Owing to the energy transfer, the amplitudes of all those space harmonics will be scaled by a common factor %themselves by interacting with the electron beam, 
and the SP radiation included in them is generated by stimulated emission. % with a high power conversion efficiency.

Notice that this scheme is different from the Orotron and SP-FEL. For the Orotron (e.g., as presented in \cite{rusin1969orotron}), the stimulated nature of the radiation from the standing wave in the cavity that directly interacts with the the electron beam is similar to this scheme. However, the structure and FP resonance mechanism of the Orotron cavity formed by two mirrors, with a grating embeded in one of them, are different from the single grating structure and BIC resonance mechanism used in this scheme. For the SP-FEL (e.g., as presented in \cite{Andrews2004}), the single grating structure used for beam-wave interaction is similar to this scheme. However, effective beam-wave interaction in the SP-FEL takes place between the electron beam and a surface mode of the grating that has no propagating space harmonic. Thus, the electron beam is bunched at the frequency of the surface mode but no SP radiation at such frequency can be enhanced. In the absence of direct interaction with the electron beam, the fields of radiative modes at the harmonics of the bunching frequency are enhanced by superradiant effect, which produces spontaneous radiation but not the stimulated radiation in this scheme~\cite{Andrews2005a}.

\section{Physics of BIC resonances} \label{sec:bic}

%The diffraction characteristics play an essential role in the scheme. Concerning a constant excitation stimulated by a continuous electron beam, a high $|r_0|$ is needed for effective beam-wave interaction. Moreover, to efficiently produce far-field radiation, high reflection coefficients of the propagating space harmonics are also favored. We will show that, making use of a strong resonance near a BIC, an enhancement of multiple space harmonics at the resonant frequency can be obtained, including both the propagating orders %for SP radiation 
%and the evanescent 0th order. % for electron bunching.
%To understand how to generate BIC resonances under evanescent wave incidence, %the field profile inside the grating should not be ignored. To this end, 
%we use the waveguide-array (WGA) modes formulation, which was introduced to explain the reflection of a plane wave impinging on a grating \cite{Chang-Hasnain2012}. Such mode-matching method not only provides a straightforward routine to extract the reflection coefficients of a grating but also leads to intuitive physical insights into its extraordinary properties \cite{doi:10.1063/1.5010144,doi:10.1063/1.5045547,Qiao:18,PhysRevLett.114.073601}. Below we generalize the analytical method and derive the WGA modes in gratings under evanescent wave incidence. %We will show that the constructive interference of WGA modes at grating interfaces results in a very strong resonance and hence the enhancement of multiple space harmonics. 

The diffraction characteristics play an essential role in the proposed scheme. Concerning a constant excitation stimulated by a continuous electron beam, high reflection coefficients of the 0th order and the propagating orders are needed for effective beam-wave interaction and efficient far-field radiation, respectively. We will show that, making use of a strong resonance near a BIC, those space harmonics can be greatly enhanced at the resonant frequency. To this end, we use the waveguide-array (WGA) modes formulation, which was initially introduced to explain the reflection of a plane wave impinging on a grating \cite{Chang-Hasnain2012}. Such mode-matching method not only provides a straightforward routine to extract the reflection coefficients of a grating but also leads to intuitive physical insights into its extraordinary properties \cite{doi:10.1063/1.5010144,doi:10.1063/1.5045547,Qiao:18,PhysRevLett.114.073601}. Below we generalize the analytical method and derive the WGA modes in gratings under evanescent wave incidence. %We will show that the constructive interference of WGA modes at grating interfaces results in a very strong resonance and hence the enhancement of multiple space harmonics.

Along the $x$ direction, the grating can be treated as a periodic array of waveguides, where a series of WGA modes exist. Concerning a TM-polarized incident wave, these WGA modes can be denoted as $\rm TM_0$, $\rm TM_1$, $\rm TM_2$, etc. The lateral magnetic field profile in the $z$ direction for the ${\rm TM}_m$ mode can then be expressed as
\begin{eqnarray}
	H_{ym}(z)&=&A_1{\rm e}^{{\rm i}k_{am}z}+A_2{\rm e}^{-{\rm i}k_{am}z}, \quad 0<z<a,
	\label{eq:field_Hy1} \\	
	H_{ym}(z)&=&B_1{\rm e}^{{\rm i}k_{bm}z}+B_2{\rm e}^{-{\rm i}k_{bm}z}, \quad a<z<p,
	\label{eq:field_Hy2}	 
\end{eqnarray}
where the $z$ wavenumbers in the low-index bar and high-index bar are $k_{am}=\left(n_a^2k_0^2-k_{xm}^2\right)^{1/2}$ and $k_{bm}=\left(n_b^2k_0^2-k_{xm}^2\right)^{1/2}$, respectively, and $k_{xm}$ is the $x$ wavenumber. Enforcing the continuity of tangential field components at the bar interfaces and the Bloch boundary condition yields the dispersion relation for WGA modes,
\begin{equation}
\eqalign{2k_{am}k_{bm}\left[\cos(k_{am}a)\cos(k_{bm}b)-\cos(k_{0}p/\beta_0)\right] \cr =(n_b^2k_{am}^2/n_a^2+n_a^2k_{bm}^2/n_b^2)\sin(k_{am}a)\sin(k_{bm}b).}
\label{eq:dispersion}
\end{equation}
\begin{figure*}[htbp!]
	\centering
	\includegraphics[width=11.0cm]{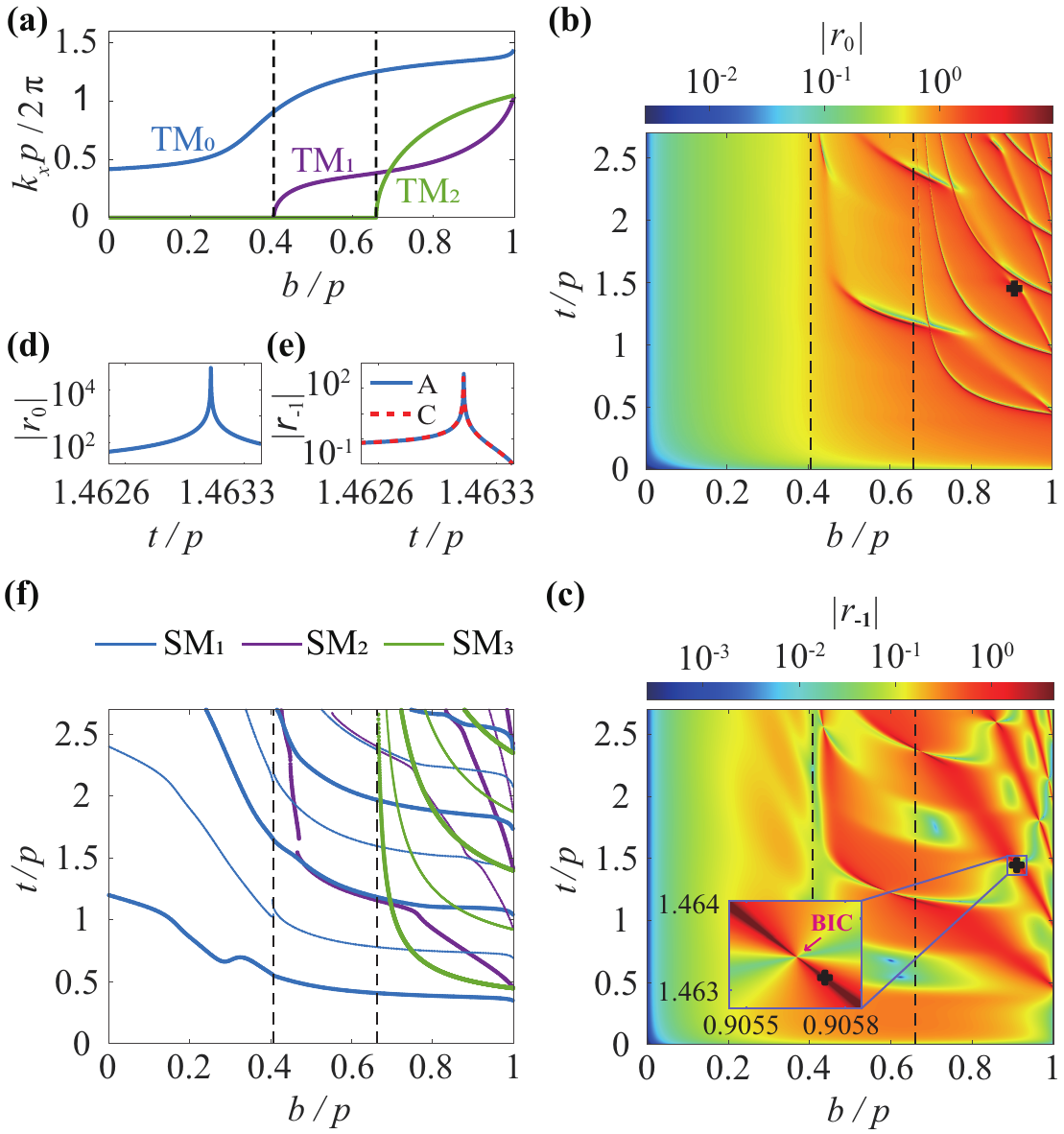}
	\caption{Strong enhancement of reflection coefficients via BIC resonances. (a) $x$ wavenumber $k_x$ of WGA modes as a function of the dutycycle $b/p$. The black dashed lines indicate the differences of three dutycycle regions: one, two, and three-mode. (b) $t-b$ map of $|r_0|$. The operation point is indicated by the black star. (c) $t-b$ map of $|r_{-1}|$. The inset shows the region near the operation point, where exists a BIC. (d) vertical slice of (b) around the operation point with $b/p=0.906$. (e) $|r_{-1}|$ calculated by the WGA analytical solution~(blue solid) and by COMSOL simulation~(red dashed) for $b/p=0.906$. (f) The combinations of thickness $t$ and dutycycle $b/p$ when the 1st-order (blue), 2nd-order (purple), and 3rd-order (green) supermodes
	%$\rm SM_1$ (blue), $\rm SM_2$ (purple), and $\rm SM_3$ (green) 
	reach their resonance conditions $\phi=2l\pi$. The thick and thin lines are for odd and even $l$ values, respectively. Parameters: electron energy $E_0=51~{\rm keV}$, grating period $p=0.416\lambda$, SP radiation angle $\theta_{-1}=90^\circ$, $n_{b}=3.486$, $n_{a}=1$.} 
	\label{fig:resonance}
\end{figure*}
Equation (\ref{eq:dispersion}) suggests the plausible control of the number of WGA modes by adjusting the dutycycle for given values of beam energy, grating period and wavelength. 
For instance, with $E_{\rm 0}=51~{\rm keV}$, $p/\lambda = \beta_0=0.416$ ($\theta_{-1}=90^\circ$), $n_{b}=3.486$, and $n_{a}=1$, we show the dependence of $k_x$ on the dutycycle $b/p$ in figure~\ref{fig:resonance}(a). As dutycycle increases, more and more modes exist. %Notice that there is a lower dutycycle limit for each mode, except for the fundamental mode. %The dual-mode region where only two modes exist is indicated by the black dash lines. 

The introduction of grating boundaries in the $x$ direction confines the WGA modes inside the grating by coupling to the outside space harmonics. Following the analytical mode-matching method proposed in \cite{Chang-Hasnain2012}, the reflection coefficients $|r_0|$ and $|r_{-1}|$ can be determined, as shown in figures~\ref{fig:resonance}(b)--\ref{fig:resonance}(e). Figure~\ref{fig:resonance}(e)
compares the values of $|r_{-1}|$ using the WGA analytical
solution and COMSOL \cite{Szczepkowicz:20}, where a satisfactory agreement is obtained. A BIC can be considered as a resonance with zero leakage and zero linewidth \cite{Hsu2016}. Figures~\ref{fig:resonance}(b)--\ref{fig:resonance}(e) reveals that, BICs can be excited for appropriate dutycycles and thicknesses. A small variation of the parameters from the BIC leads to a strong radiative resonance %(BIC resonance) 
such as the operation point \cite{Yang2018}, where both $|r_0|$ and $|r_{-1}|$ are greatly enhanced. 

The highly-ordered patterns in figures~\ref{fig:resonance}(b) and \ref{fig:resonance}(c) reveal %the strong dependence of reflection coefficients on both dutycycle and grating thickness, indicating 
an interference effect. We will show that this phenomenon, as well as the origin of BIC resonances, are linked with the FP-resonance mechanism of those WGA modes. Inside the grating, the magnetic field including a number of WGA modes propagating downward can be given by 
\begin{equation}
	{H}_y(x,z) = [H_{y0}(z)~H_{y1}(z)~H_{y2}(z) \dots]{\rm e}^{{\rm i}{\bi k}_xx}[C_0~C_1~C_2 \dots]^T,
	\label{eq:WGAfields}
\end{equation}
where ${\bi k}_x$ is a diagonal propagation matrix composed of $k_{xm}$, and ${\bi C}=[C_0~C_1~C_2 \dots]^T$ is a state vector\cite{Qiao:18}. %characterizes the field \cite{Qiao:18}. 
After a round trip in the grating, the state vector becomes $\bi{MC}$, where $\bi{M}= {\bi R}{\rm e}^{{\rm i}{\bi k}_xt}{\bi R}{\rm e}^{{\rm i}{\bi k}_xt}$ is the propagation matrix, and ${\bi R}$ the reflection matrix at the interfaces. The original $\bi{M}$ obtained by the mode‐matching method is undiagonal, i.e., WGA modes are not orthogonal. The diagonalization of the $\bi{M}$ matrix provides a new set of orthogonal modes, namely ``supermodes'' \cite{Chang-Hasnain2012}. Accordingly, the phase shift accumulated by the $m$th-order supermode (${\rm SM}_{m}$) after a round trip corresponds to the phase of the $m$th eigenvalue of ${\bi M}$, denoted by $\phi_m$. FP resonances occur when the round-trip phase shift $\phi_m$ is $2l\pi$, where $l$ is an integer. 

%In figure~\ref{fig:resonance}(f), the combinations of $t$ and $b/p$ supporting the resonance dominated by the 1st, 2nd, 3rd-order supermode (denoted $\rm SM_1$, $\rm SM_2$, $\rm SM_3$) are shown by the blue, purple, and green curves, respectively, and the thick and thin lines illustrate the value of $l$ being odd and even, respectively.
Figure~\ref{fig:resonance}(f) shows the combinations of thickness %$t$ 
and 
ducycycle %$b/p$ 
supporting the FP resonances.
The patterns in the $|r_0|$ and $|r_{-1}|$ maps are grided by different resonance curves, which indicates different contributions to the space harmonics from each supermode. A strong BIC resonance can be obtained when ${\rm SM}_{1}$ and ${\rm SM}_{2}$ reach their conditions in phase, with $|\phi_{1}-\phi_{2}|=2s\pi$, $s$ being an even integer. %In this case, both $|r_0|$ and $|r_{-1}|$ are simultaneously enhanced. 
We will show that stimulated SP radiation can be eventually enabled in the presence of the interaction between an electron beam and the BIC resonance.

\begin{figure}[bp!]
		\centering
		\includegraphics[width=12cm]{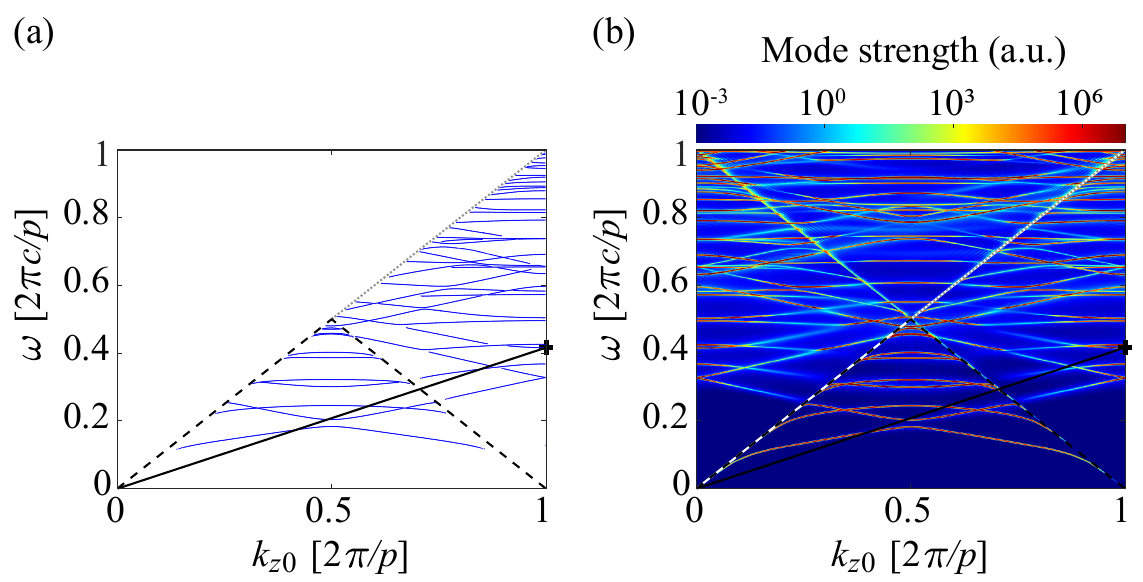}
		\caption{Dispersion curves calculated with WGA analytical solution (a) and FDTD (b) of a grating with $t=1.463p$ and $b=0.906p$. The beam line of energy 51~keV is marked by the black solid line, and the black star indicates the operation point. The region of surface modes is located below the black dashed lines. }
		\label{fig:band1}
	\end{figure}

Note that the supermode being self-sustainable requires ${\bi{(I-M)C}}=0$~\cite{Chang-Hasnain2012}. For a nontrivial solution to exist, the determinant of the coefficients should be zero, yielding the dispersion relation:
		\begin{eqnarray}
			{\rm{det}}\bi {(I-M)}=0, 
			\label{eq:disp}
		\end{eqnarray}
		where ${\bi{I}}$ is an identity matrix. %Equation~\ref{eq:disp} describes the resonance condition of the internal supermode, which can also be derived using the above mode-matching method in the absence of incident field (see supplementary material for the formulation). 
		%To study the beam-wave interaction, 
		The combinations of $k_{z0}$ and $\omega$ corresponding to the minima of ${\rm{det}}\bi {(I-M)}$ being less than 0.5 are shown in figure~\ref{fig:band1}(a) for a grating with $t=1.463p$ and $b=0.906p$, which are the same as the operation point in figure~\ref{fig:resonance}. Here, only the reflection coefficients below the light line ($\omega/k_{z0}=c$) are calculated, because the excitation is stimulated by an evanescent incident wave. Figure~\ref{fig:band1}(b) shows a full dispersion diagram obtained by Lumerical FDTD simulation with Bloch boundary conditions on a single grating period. Good agreement between the analytical solutions and the FDTD simulation results below the light line indicates the correctness of the dispersion relation given by equation (\ref{eq:disp}).
	
Interestingly, the analytical method introduced here is readily to be used for the inverse SP effect. Dielectric laser accelerators based on such effect have shown great potentials in recently years~\cite{Sapra79}. Upon the incidence of a laser on the grating, one synchronous evanescent space harmonic generated by the diffraction effect can efficiently accelerate electrons~\cite{RevModPhys861337}. Compared to the SP effect with evanescent wave incidence, the analysis for the inverse SP effect with propagating wave incidence follows roughly the same steps, except that the expression in (\ref{eq:field_in}) should be modified to match the incident laser field. The accelerating gradient depends on the calculated reflection or transmission coefficients for the synchronous space harmonic, which may also be enhanced by the BIC resonances. If the laser wavelength, incident angle and electron energy of the inverse SP effect agree with the radiation wavelength, radiation angle and electron energy of the SP effect, the required grating parameters to enable BICs should be the same.

\section{Beam-wave interaction}

\begin{figure}[bp!]
	\centering
	\includegraphics[width=12cm]{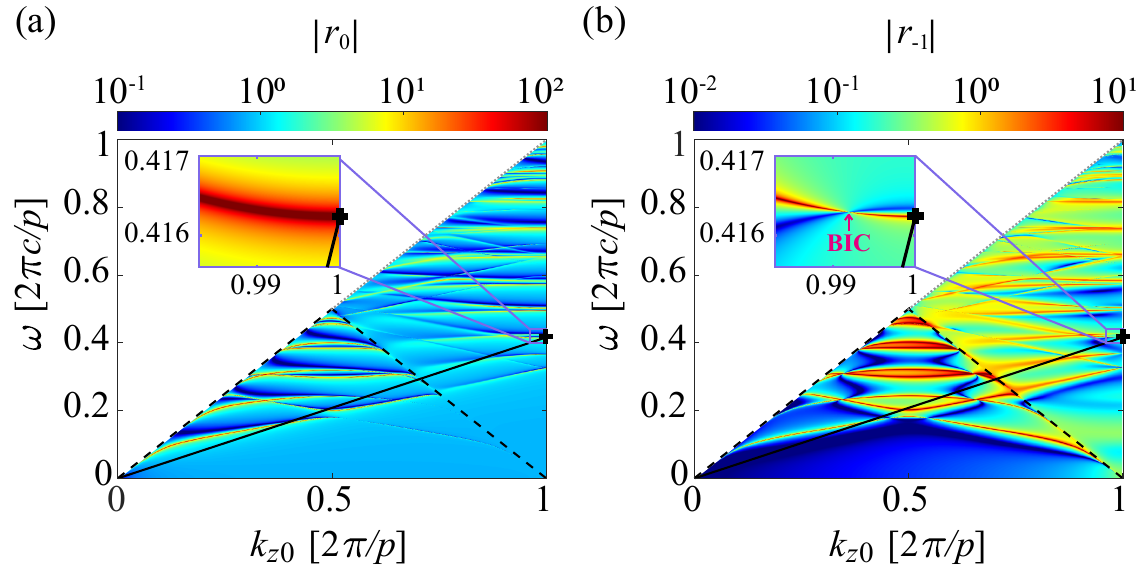}
	\caption{Reflection coefficients $|r_0|$ (a) and $|r_{-1}|$ (b) of a grating with $t=1.463p$ and $b=0.906p$. The beam line of energy 51~keV is marked by the black solid line, and the black star indicates the operation point. The region of surface modes is located below the black dashed lines. The insets show the region near the operation point, where exists a BIC.} 
	\label{fig:band}
\end{figure}

Firstly, we consider the interaction between a 51~keV electron beam and a grating by following the parameters of the operation point in figure~\ref{fig:resonance}. To study the beam-wave interaction, the reflection coefficient $|r_0|$ and $|r_{-1}|$ for a grating with $t=1.463p$ and $b=0.906p$ are also calculated by the WGA analytical solution versus angular frequency $\omega$ and $k_{z0}$, as shown in figure~\ref{fig:band}. These diagrams show the same dispersion curves as in figure~\ref{fig:band1}, with different dispersion curves illustrated by extraordinarily high reflection coefficients~\cite{Qiao:18}. The operation point marked by the black star, which is also shown in figure~\ref{fig:band1}, is the intersection of the 51~keV beam line with one dispersion curve. %The region of surface modes is located below the black dashed lines in figure~\ref{fig:band}. 
The operation point is located above the region of surface modes, corresponding to a radiative mode with the {-1st}-order space harmonic being the SP radiation. A strong resonance is induced at the operation point by a nearby BIC for effective beam-wave interaction, as depicted in the insets of figure~\ref{fig:band}(b).

Now we show that the beam-wave interaction enables electron bunching and thus stimulated SP radiation through two-dimensional simulations with a finite-difference-time-domain-based fully electromagnetic code CHIPIC \cite{5233920,Liang2017,Liu2018a,zhou2016electron}. While our theory is general, we choose an operation frequency of 0.2~THz to exemplify our concept. Grating and electron parameters are determined by the operation point in figure~\ref{fig:band}. A grating with dutycycle $b/p=0.906$, thickness $t=913~{\rm \mu m}$, period $p=624~{\rm \mu m}$ and number of periods $N=30$ is used. On top of the grating, a sheet electron beam of energy $51~{\rm keV}$, thickness ${\rm 100~\mu m}$, and current density ${\rm 50~A/cm^2}$ are placed in an external uniform longitudinal magnetic field of 0.7~${\rm T}$. %Electron and grating parameters are in accordance with the operation point in figure~\ref{fig:band}. %, with SP radiation along the normal direction. %Moreover, the following parameters have been used, ...
%Consider a grating of period $p=624~{\rm \mu m}$, dutycycle $b/p=0.906$, thickness $t=913~{\rm \mu m}$ and number of periods $N=30$, all placed in an external uniform longitudinal magnetic field of $0.5~{\rm T}$. On top of the grating, a sheet electron beam of energy $51~{\rm keV}$, thickness ${\rm 100~\mu m}$, and current density ${\rm 50~A/cm^2}$ are placed. 
Considering the small transverse extent of the 0th-order space harmonic, the spacing between the beam and grating %between the electron beam and the grating 
is set to be 50~${\rm \mu m}$. We note that such a single-layer grating could be realized exploiting lithographic and etching techniques \cite{Wagner2010}, and the required electron density can be achieved with state-of-the-art emitters \cite{7323834}.

\begin{figure}[bp!]
	\centering
	\includegraphics[width=15cm]{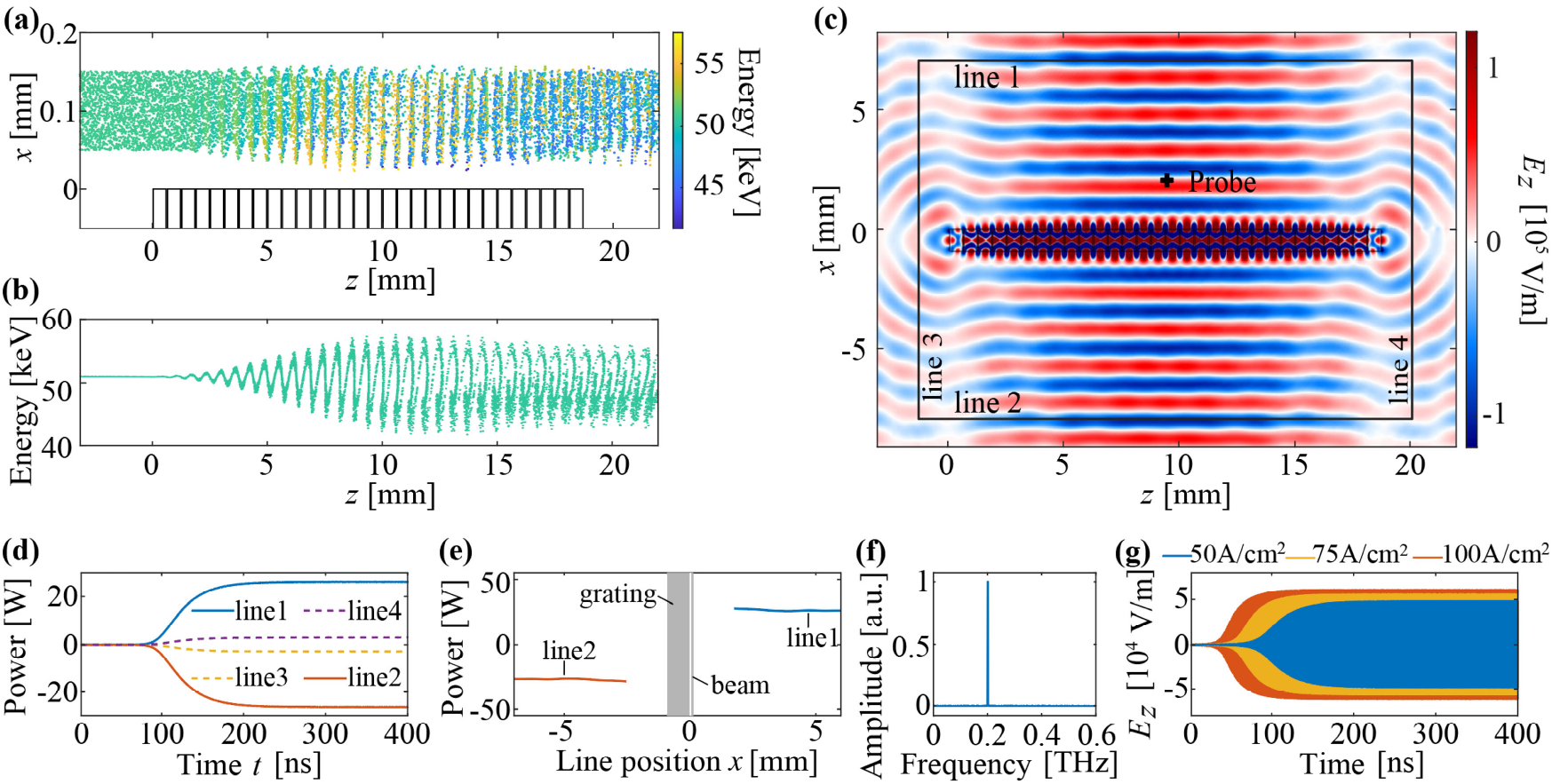}
	\caption{Simulation results of coherent SP radiation source based on BIC resonances, with electron energy $E_0=51~{\rm keV}$. (a) Snapshot of the particle distribution above the grating. (b) Phase-space distribution of the electron beam. (c) Field pattern of the $E_z$ component.(d) Time dependence of the upward power through line 1 and line 2, and the power towards right through line 3 and line 4. The line locations are shown in (c). (e) Upward power recorded by line 1 and line 2 at different $x$ positions, with the simulation time $t=300~{\rm{ns}}$. (f) Field spectrum of the SP radiation at the point marked by the black star in (c). (g) $E_z$ fields as a function of time at the point marked by a black star in (c), with current density $J_z=50~{\rm{A/cm^2}}, 75~{\rm{A/cm^2}}, 100~{\rm{A/cm^2}}$. The current density in (a--f) is $50~{\rm{A/cm^2}}$.} %Parameters: grating period $p=624~{\rm \mu m}$, grating dutycycle $b/p = 0.906$, grating thickness $t=913~{\rm \mu m}$, beam thickness $100~{\rm \mu m}$, the numbers of grating period $N=30$.} 
	\label{fig:simulation}
\end{figure}

Figures~\ref{fig:simulation}(a) and \ref{fig:simulation}(b) show the particle and phase-space distributions of the electron beam, respectively. 
Along the longitudinal direction, the electrons interact with the self-excited resonant radiative mode, inducing an energy modulation. After a distance, %as shown by the inset in figure~\ref{fig:simulation}(a), 
faster electrons intersect slower ones, so the energy modulation develops into density modulation. Owing to the periodicity of the 0th-order space harmonic along the $z$ direction ($k_{z0}p=2\pi$), electron bunches are separated by one grating period. Due to the evanescent nature of the 0th-order space harmonic, electrons closer to the grating experience stronger energy modulation and therefore form bunches earlier. On the right side of figure~\ref{fig:simulation}(b), there is a significant decrease in the overall electron energy, resulting from the beam-to-wave energy conversion.

Figures~\ref{fig:simulation}(c) shows the $E_z$ field pattern, indicating that the SP radiation is indeed propagating in the normal direction ($\theta=90^\circ$). It should be noted that the finite grating length in the $z$ direction leads to omnidirectional radiation at both ends. To evaluate the efficiency, the radiation powers towards different directions are detected by four lines shown in figure~\ref{fig:simulation}(c). Assuming a grating width of 1~mm, figure~\ref{fig:simulation}(d) indicates that, the maximum total power of the SP radiation propagating in the normal direction (the sum of the powers through line 1 and line 2) is about 52.8~W, and the maximum total power of the radiation towards other directions (the sum of the powers through line 3 and line 4) is about 5.9~W. In comparison, the power in the normal direction is much higher than the power towards other directions. The corresponding power conversion efficiency for the SP radiation through line 1 and line 2 is about 2.1\%. In figure~\ref{fig:simulation}(e), by changing the $x$ positions of line 1 and line 2, we found that the radiation power keeps almost constant in the normal direction, indicating the SP radiation is fairly directional. Figure~\ref{fig:simulation}(f) shows that the radiation frequency detected by the probe marked by a black star in figure~\ref{fig:simulation}(c) is about 0.2~THz, consistent with the prediction by (\ref{eq:field_angle}). Figure~\ref{fig:simulation}(g) shows the $E_z$ fields detected by the probe for different current densities, indicating that the radiation can be potentially enhanced by increasing the electron beam current.

\begin{figure}[htbp]
	\centering
	\includegraphics[width=12cm]{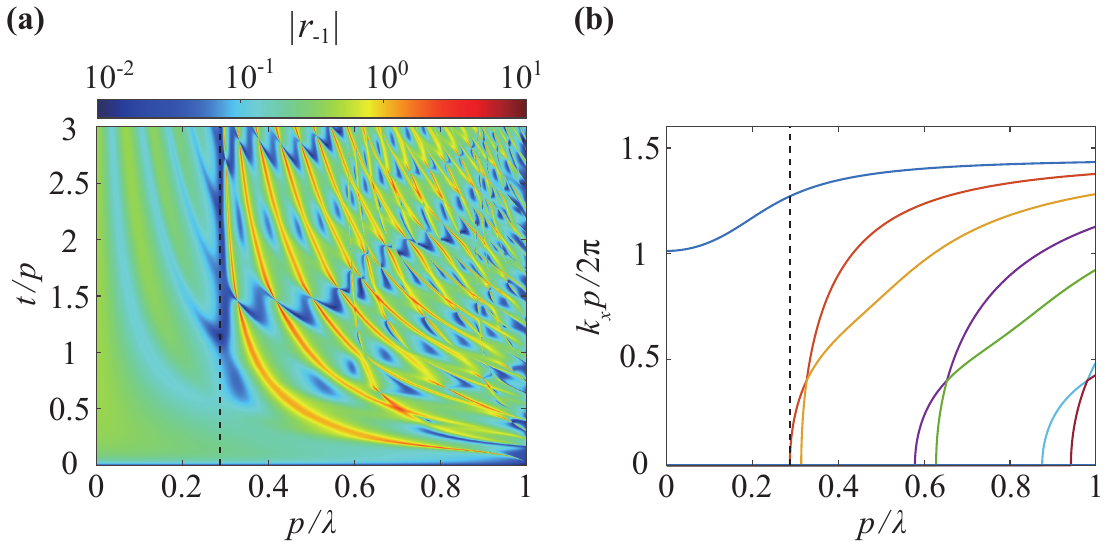}
	\caption{(a) Reflection coefficient $|r_{-1}|$ as a function of grating period $p$ and thickness $t$. (b) $x$ wavenumber $k_x$ of WGA modes as a function of $p$. The black dashed line indicates the inferior limit of $p/\lambda$ enabling at least two WGA modes. Parameters: radiation frequency 0.2~THz, radiation angle $\theta_{-1}=90^\circ$, grating dutycycle $b/p=0.906$.}
	\label{fig:limit}
\end{figure}

Secondly, we consider the interaction with a smaller electron energy, which would be desirable for a compact and cost-effective free-electron radiation source. When the operation frequency and radiation angle are held constant, it is possible to use a smaller electron energy by reducing the grating period, as shown in figure~\ref{fig:limit}(a). For example, with grating period $p=492~{\rm{\mu m}}$, the required electron energy to radiate the -1st-order space harmonic at $\theta_{-1}=90^\circ$ is reduced to 30~keV. However, as shown in figure~\ref{fig:limit}(b), there is an inferior limit for the electron's normalized velocity $\beta_0=p/\lambda$ to support at least two WGA modes propagating inside the grating, which is required to enable BICs. To solve this problem, we can use higher-order space harmonics as SP radiation.

\begin{figure}[bp]
	\centering
	\includegraphics[width=12cm]{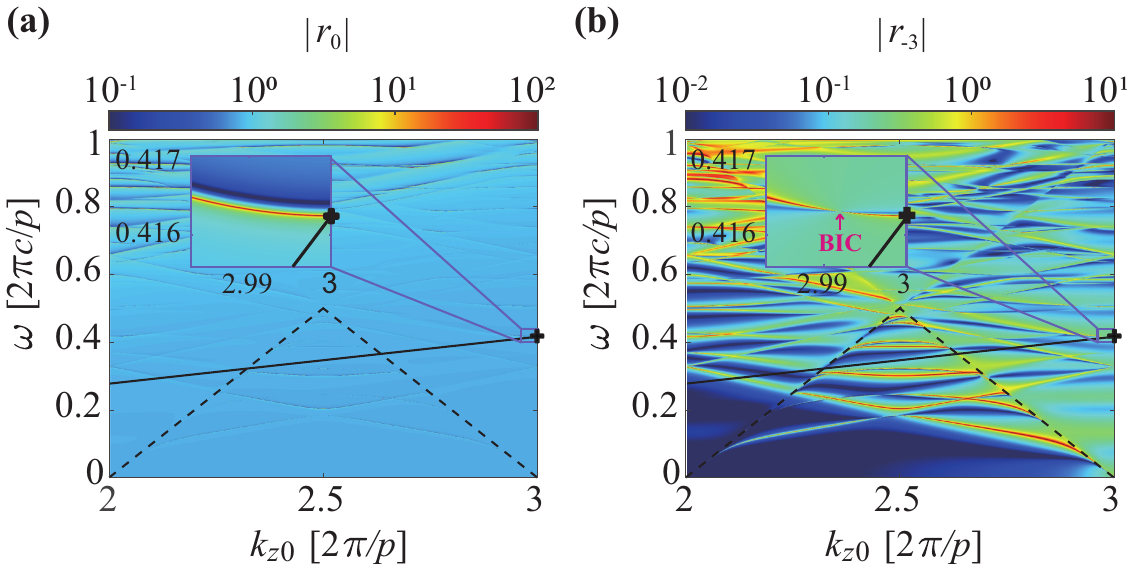}
	\caption{Reflection coefficient $|r_0|$ (a) and $|r_{-3}|$ (b) of a grating with $t=1.463p$ and $b=0.906p$. The beam line of energy 4.99~keV is marked by the solid black line, and the black star indicates the operation point. The region of surface modes is located below the black dashed lines. } 
	\label{fig:band2}
\end{figure}

%By utilizing a higher-order space harmonics for SP radiation, it is possible to apply our concept for developing low-voltage radiation sources. 
To exemplify the method, the grating period $p$, dutycyle $b/p$, thickness $t$ and the radiation wavelength $\lambda$ remain the same as in the previous discussions. For 51~keV electrons, we have $k_0p/\beta_0=2\pi$. %Equation (\ref{eq:dispersion}) indicates that, 
When changing the normalized electron velocity to $\beta_0/j$ ($j=2,3,4,{\rm etc.}$), the solution of the dispersion relation for WGA modes stays the same, and thus the operation frequencies of those BIC resonances remain unchanged. %Meanwhile, the $-j$th space harmonic with $k_{z(-j)}=0$ becomes propagating along the surface normal direction. 
%In this case, using the BIC-enhanced $-j$th space harmonic for SP radiation and the BIC-enhanced 0th for bunching allows us to obtain coherent SP radiation with lower electron energies. 
In the following example, we set $j=3$, corresponding to an electron energy of 4.99~keV. Figure~\ref{fig:band2} shows the corresponding reflection coefficients $|r_0|$ and $|r_{-3}|$, where $k_{z0}$ ranges from $4\pi/p$ to $6\pi/p$. Compared with the 51 keV case, the locations of the dispersion curves %indicated by high reflection coefficients 
remain the same, but the line widths become narrower. The operation point is then determined by the intersection of the 4.99 keV beam line with one dispersion curve, corresponding to a radiative mode with the -3rd-order space harmonic being the SP radiation. It is still located near a BIC and thus an enhancement of both $r_0$ and $r_{-3}$ can be obtained. %Keeping the grating parameters unchanged, the values of $|r_0|$ and $|r_{-3}|$ are plotted in figure~\ref{fig:band2} versus $\omega$ and $k_{z0}$, where $k_{z0}$ ranges from $4\pi/p$ to $6\pi/p$. Compared with figure~\ref{fig:band}, the locations of the dispersion curves indicated by high reflection coefficients remain the same, but the line widths become narrower. The operation point indicated by the black star, which is the intersection of the 4.99~keV beam line with one dispersion curve, corresponds to a radiative mode with the -3rd-order space harmonic being the SP radiation. It is still located near a BIC and thus an enhancement of both $r_0$ and $r_{-3}$ at the operation point can be obtained.

%In the following example, we set $j=3$, with electron energy $E_0=4.99~{\rm keV}$. Figure~\ref{fig:band2} shows the corresponding reflection coefficients $|r_0|$ and $|r_{-3}|$. Compared with the 51~keV case, the locations of the dispersion curves indicated by high reflection coefficients remain the same, but the line widths become narrower. The operation point is then determined by the intersection of the 4.99~keV beam line with one dispersion curve, corresponding to a radiative mode with the -3rd-order space harmonic being the SP radiation. It is still located near a BIC and thus an enhancement of both $r_0$ and $r_{-3}$ can be obtained. %Keeping the grating parameters unchanged, the values of $|r_0|$ and $|r_{-3}|$ are plotted in figure~\ref{fig:band2} versus $\omega$ and $k_{z0}$, where $k_{z0}$ ranges from $4\pi/p$ to $6\pi/p$. Compared with figure~\ref{fig:band}, the locations of the dispersion curves indicated by high reflection coefficients remain the same, but the line widths become narrower. The operation point indicated by the black star, which is the intersection of the 4.99~keV beam line with one dispersion curve, corresponds to a radiative mode with the -3rd-order space harmonic being the SP radiation. It is still located near a BIC and thus an enhancement of both $r_0$ and $r_{-3}$ at the operation point can be obtained.

\begin{figure}[tbp!]
	\centering
	\includegraphics[width=15cm]{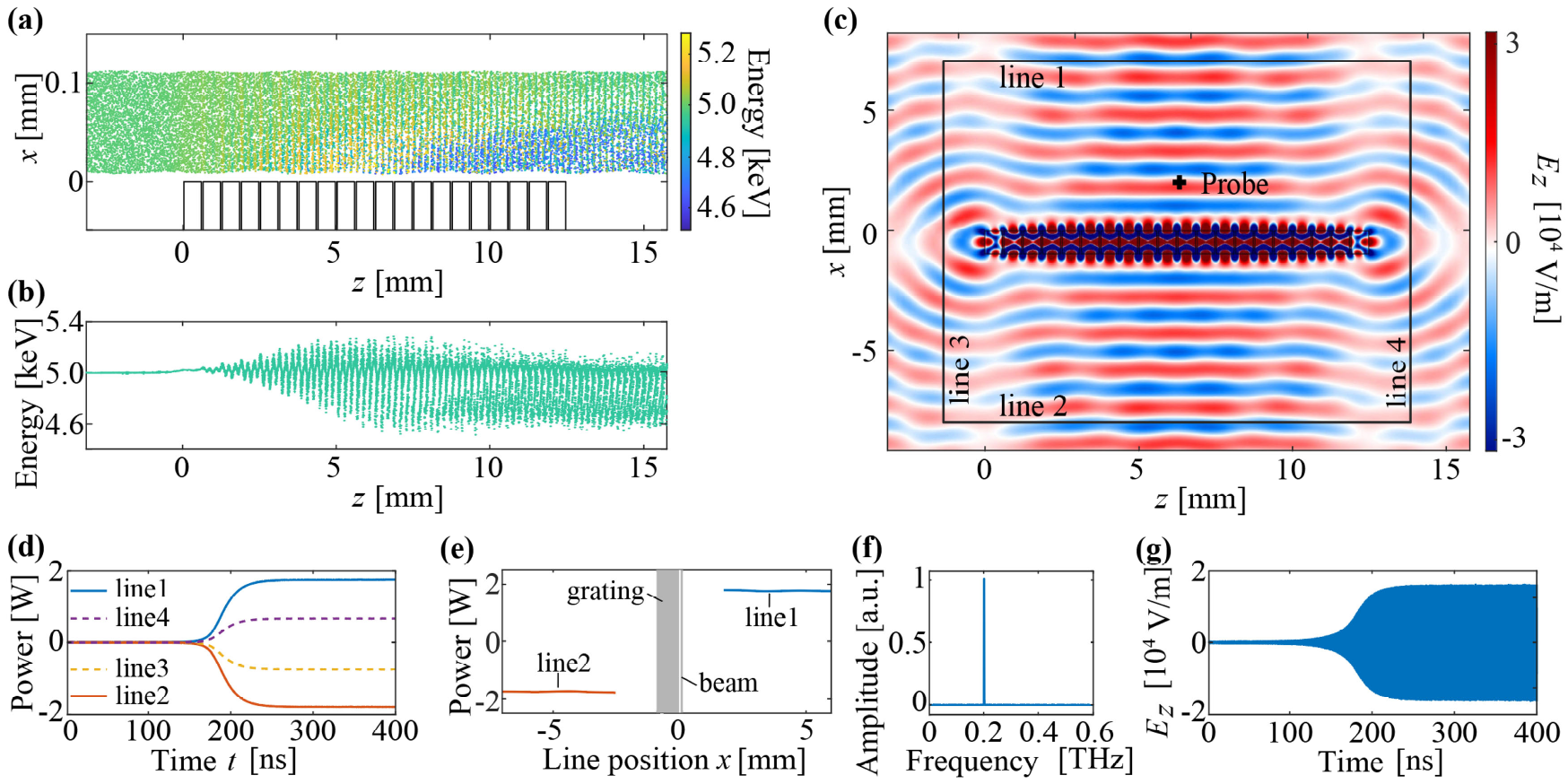}
	\caption{Simulation results of low-voltage coherent SP radiation sources based on BIC resonances, with electron energy $E_0=4.99~{\rm keV}$ and current density $J_z=50~{\rm{A/cm^2}}$. (a) Snapshot of the particle distribution above the grating. (b) Phase-space distribution of the electron beam. (c) Field pattern of the $E_z$ component. (d) Time dependence of the upward power through line 1 and line 2, and the power towards right through line 3 and line 4, with the line locations being shown in (c). (e) Upward power detected by line 1 and line 2 at different $x$ positions, with the simulation time $t=300~{\rm{ns}}$. Spectrum (f) and the $E_z$ fields (g) of the SP radiation at the point marked by a black star in (c).}%Parameters: grating period $p=624~{\rm \mu m}$, grating dutycycle $b/p = 0.906$, grating thickness $t=913~{\rm \mu m}$, beam thickness $100~{\rm \mu m}$, the numbers of grating period $N=20$.} 
	\label{fig:simulation2}
\end{figure}

Figures~\ref{fig:simulation2} shows the particle simiulation results for 4.99~keV electrons, with the number of grating period $N=20$, the spacing between the beam and grating $10~{\rm{\mu m}}$, the current density $50~{\rm{A/cm^2}}$ and the other parameters the same as figure~\ref{fig:simulation} . 
%Comparing with figures~\ref{fig:simulation}(a) and \ref{fig:simulation}(b), 
Comparing with the 51~keV case, the spacing between electron bunches reduces to $p/3$, simply because $k_{z0}=6\pi/p$, as shown in figures~\ref{fig:simulation2}(a) and \ref{fig:simulation2}(b). Figures~\ref{fig:simulation2}(c) and \ref{fig:simulation2}(f) demonstrate that the SP radiation angle and frequency remain the same as in the 51~keV case. Figure~\ref{fig:simulation2}(d) and \ref{fig:simulation2}(e) show the maximum total power for the SP radiation through line 1 and line 2 is about 3.6~W, implying an efficiency of 1.4\%, which is lower than that for 51~keV electrons owing to the different reflection coefficients and the smaller transverse extent of the 0th-order space harmonic. We note that in this case the fraction of the power through line 3 and line 4 is relatively high (about 1.3~W), this might be caused by the fact that the number of periods becomes less, leading to a lower quality factor for the grating resonator~\cite{taghizadeh2017quasi}. Figure~\ref{fig:simulation2}(g) shows that the $E_z$ field of SP radiation detected by the probe in figure~\ref{fig:simulation2}(c) also becomes lower.
%Compared with using  $E_0=51~{\rm keV}$, the decreasing of power conversion efficiency results from the differences in reflection coefficients and the smaller transverse extent of the 0th diffraction order. %This example is also a piece of clear evidence to show that the emitted wave is SP radiation rather than Cherenkov radiation.

\begin{figure}[tbp!]
	\centering
	\includegraphics[width=15cm]{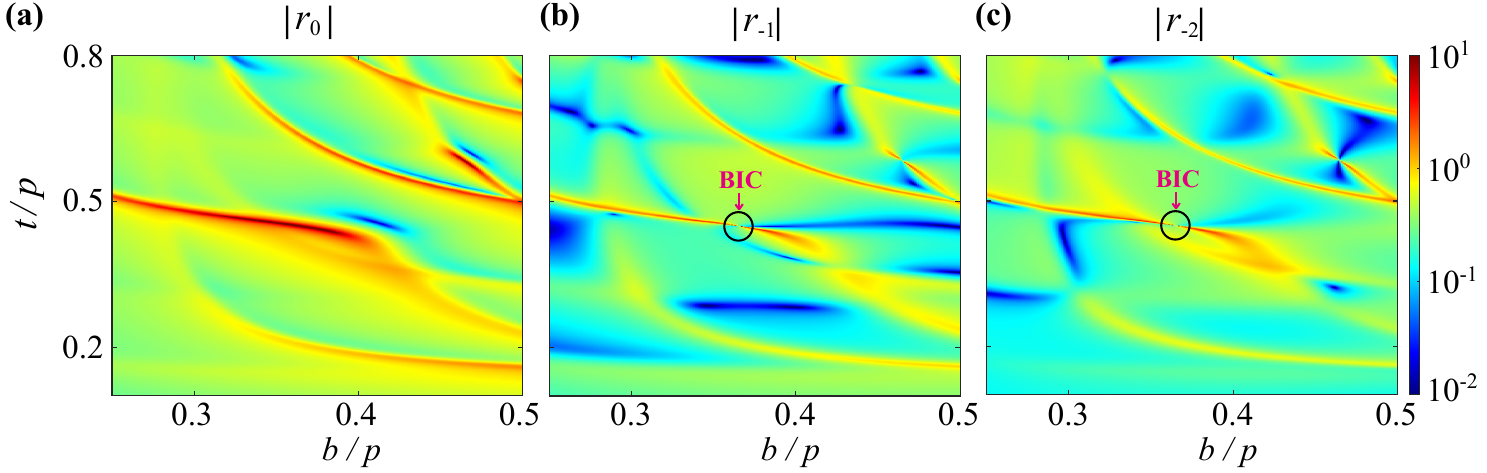}
	\caption{Reflection coefficients $|r_0|$ (a), $|r_{-1}|$ (b) and $|r_{-2}|$ (c) of a grating versus grating dutycycle $b/p$ and thickness $t$. with period $p=1499~{\rm{\mu m}}$ and electron energy $E_0=174.6~{\rm{keV}}$. The black circle in (b) and (c) indicates the location of a BIC.}
	\label{fig:simulation3}
\end{figure}

Thirdly, we consider the interaction with a radiative mode that has more than one propagating space harmonic. Based on (\ref{eq:field_angle}), with the target operation frequency remaining 0.2~THz, the electron energy and the grating period are changed to 174.6~keV and 1499~${\rm{\mu m}}$, respectively, so that the -1st-order space harmonic can radiate at $\theta_{-1}=60^\circ$ and -2nd-order at $\theta_{-2}=120^\circ$. Figure~\ref{fig:simulation3} plots the reflection coefficients $|r_0|$, $|r_{-1}|$, $|r_{-2}|$ as a function of grating dutycycle $b/p$ and thickness $t$. BICs can be found at the same locations in the contours of $|r_{-1}|$ and $|r_{-2}|$. In the following, the grating dutycycle is set to be $b/p=0.367$ and the thickness $t=0.448p$, corresponding to a radiative resonance near the BIC located inside the black circle in figures~\ref{fig:simulation3}(b) and \ref{fig:simulation3}(c).

\begin{figure}[tbp!]
	\centering
	\includegraphics[width=15cm]{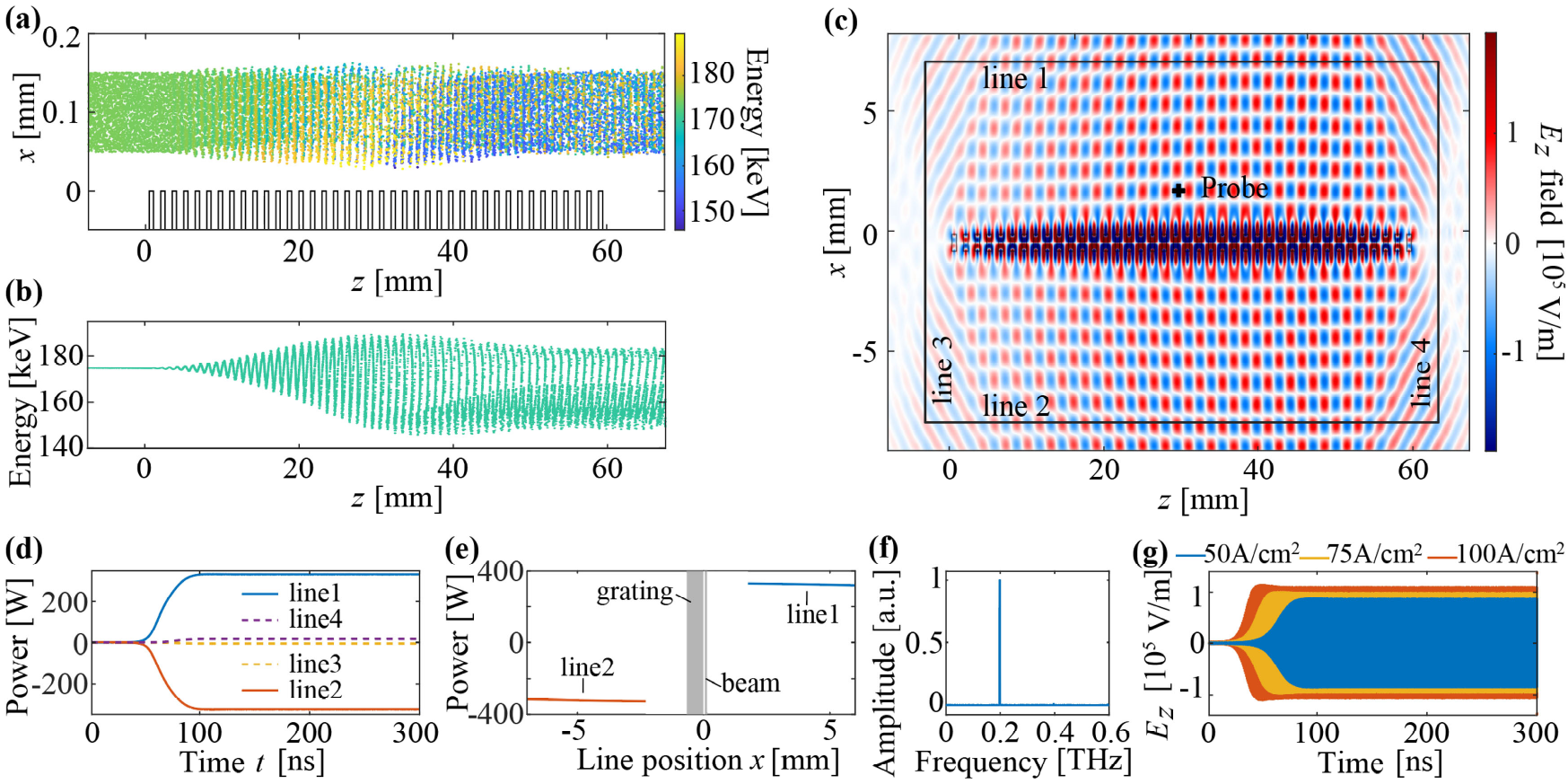}
	\caption{Simulation results of coherent SP radiation source emitting two diffraction orders, with electron energy $E_0=174.6~{\rm keV}$. (a) Snapshot of the particle distribution above the grating. (b) Phase-space distribution of the electron beam. (c) Field pattern of the $E_z$ component. (d) Time dependence of the upward power through line 1 and line 2, and the power towards right through line 3 and line 4, with the line locations being shown in (c). (e) Upward power detected by line 1 and line 2 at different $x$ positions, with the simulation time $t=200~{\rm{ns}}$. (f) Spectrum  of the SP radiation at the point marked by a black star in (c). (g) $E_z$ fields at the point marked by a black star in (c), with current density $J_z=50~{\rm{A/cm^2}}$, $75~{\rm{A/cm^2}}$, $100~{\rm{A/cm^2}}$. The current density in (a--f) is $50~{\rm{A/cm^2}}$.}
	\label{fig:simulation5}
\end{figure}

Figure~\ref{fig:simulation5} shows the particle simulation results, with the spacing between the beam and grating being $50~{\rm{\mu m}}$, and the number of grating period $N=40$. With $k_{z0}=3\pi/p$, the spacing between electron bunches is $2p/3$, as shown in figures~\ref{fig:simulation5}(a) and \ref{fig:simulation5}(b). Figures~\ref{fig:simulation5}(c) and \ref{fig:simulation5}(f) demonstrate that the SP radiation indeed includes two diffraction orders (one propagates at $\theta=60^\circ$ and the other at $\theta=120^\circ$), and the radiation frequency is about 0.2~THz. Figure~\ref{fig:simulation5}(d) shows that the maximum total power (667~W, corresponding to an efficiency of 7.6\%) through line 1 and line 2 in figure~\ref{fig:simulation5}(c) is much higher than the total power (23~W) through line 3 and line 4. Because of the oblique radiation angle, the detected power becomes less when line 1 and line 2 move away from the grating, as shown in figure~\ref{fig:simulation5}(e). Figures~\ref{fig:simulation5}(g) shows that the $E_z$ field of SP radiation detected by the probe in figure~\ref{fig:simulation5}(c) becomes stronger with an increasing electron beam current.

In these particle simulations, special attention should be paid to the electrons' lateral displacement in the $x$ direction. For example, in figure~\ref{fig:simulation5}(a), the overall beam thickness becomes significantly larger when the electrons start to bunch. This phenomenon is caused by the combined effect of the increasing space-charge repulsive forces between bunched electrons and the lateral deflection force exerted by the $E_x$ and $H_y$ fields of the synchronous 0th-order space harmonic\cite{RevModPhys861337}. Such lateral displacement may lead to charging and overheating of the grating, which is undesired for dielectric structures, especially when the spacing between the beam and grating is small. For practical implementation, to sustain the sheet electron beam near the grating, a strong longitudinal magnetic field can be used for transversely confining the beam. 

Furthermore, we note that the saturation time of the radiation in figures~\ref{fig:simulation}, \ref{fig:simulation2} and \ref{fig:simulation5} are different. The growth rate of the beam-wave interaction is the key parameter that determines the saturation time, which depends not only on the grating structures but also on the electron beam parameters\cite{Andrews2004}. Generally, enhancing the interaction can improve the growth rate and shorten the saturation time\cite{doi:10.1063/1.4711803}. % For example, compared with figures~\ref{fig:simulation}, the higher electron energy used in figure~\ref{fig:simulation5} corresponds to a 0th-order space harmonic decaying slower in the $x$ direction, which is beneficial for shortening the saturation time.
Although the analysis of growth rate goes beyond the WGA model in this article, a simple method for shortening the saturation time has been revealed by the particle simulation, i.e., increasing the current density as shown in figures~\ref{fig:simulation}(g) and \ref{fig:simulation5}(g)~\cite{cao2015Enhance}.

\section{Conclusion}

We have investigated a radiating source using a BIC resonance of a single open grating to interact with a continuous electron beam.Theoretical analysis using the mode-matching method shows that the BIC resonance can be obtained by engineering the interference of WGA modes. Particle simulations indicate the feasibility of producing coherent SP radiation in the presence of interaction between a continuous electron beam and the BIC resonance. For an incident electron energy of 51~keV, the output power of the stimulated SP radiation in the normal direction is 52.8~W with a power conversion efficiency of 2.1~\%, promising for practical applications. Employing the 3rd-order space harmonic for SP radiation, the required electron energy can be reduced to 4.99~keV. Moreover, stimulated radiations towards different directions are shown to be possible by utilizing a radiative mode that has more than one propagating space harmonic. This study shows a viable solution towards a low-voltage stimulated radiation source with the merit of simplicity. Extra optimizations of material and structure may further improve the radiation performances and would be considered for future studies.
%a strong BIC resonance could be obtained in a grating by engineering the interference of WGA modes. The interaction between a continuous electron beam and the BIC resonance of a grating can lead to coherent oscillation, enabling stimulated SP radiation. By using a higher-order space harmonic for SP radiation, the required electron energy can be reduced. 
%As an outlook, we note that our study suggests that similar devices could be realized with different materials and structures by making use of the strong resonances near BICs. Our study may enable a viable way of realizing a low-voltage coherent radiation source with a simple structure.

\section*{Acknowledgments}
This work was supported by ``the Fundamental Research Funds for the Central Universities'' (No.~2242020K40022).

%\normalem

\section*{References}
\bibliography{Main}% Produces the bibliography via BibTeX.

\end{document}